\documentclass[aps,prd,onecolumn,superscriptaddress]{revtex4-2}
\usepackage{amsmath,amssymb,bm,physics,graphicx,mathtools,cancel,bbm,color}
\usepackage{multirow,makecell}
\usepackage[T1]{fontenc}
\usepackage{lmodern}
\usepackage{float}
\newcommand{\M}{{\mathsf{M}}}
\usepackage{hyperref}
\hypersetup{
  pdftitle={Flavor Symmetry Breaking and the Fate of the First-Order Chaos-Integrability Transition in Multi-Flavor Double-Scaled SYK},
  pdfauthor={Hao Dai, Yu-Ge Chen, Li-Guo Qin, Li-Jun Tian}
}

\begin{document}

\title{Flavor Symmetry Breaking and the Fate of the First-Order
Chaos--Integrability Transition in Multi-Flavor Double-Scaled SYK}

\author{Hao Dai}
\email{Isaac.H.Dai@outlook.com}
\affiliation{Department of Physics, Shanghai University, Shanghai 200444, China}
\author{Yu-Ge Chen}
\email{chenyuge@shu.edu.cn}
\affiliation{Institute for Quantum Science and Technology, Shanghai University,
Shanghai 200444, China}
\author{Li-Guo Qin}
\email{lgqin@foxmail.com}
\affiliation{School of Mathematics, Physics and Statistics, Shanghai University
of Engineering Science, Shanghai 201620, China}
\author{Li-Jun Tian}
\email{tianlijun@shu.edu.cn}
\thanks{Corresponding author.}
\affiliation{Department of Physics, Shanghai University, Shanghai 200444, China}

\begin{abstract}
We study the fate of the first-order chaos--integrability transition of the
Berkooz--Brukner--Jia--Mamroud (BBJM) model when the integrable sector is
replaced by $d$ commuting-SYK flavors that do not commute with one another,
i.e.\ the $d$-commuting construction of Gao, Lin and Peng. A natural large-$d$ treatment, using the average $\bar q(d)$ of Gao, Lin and
Peng in the two-chord model of Berkooz et al., suggests that the transition
melts already at $d=2$.
Solving the color-resolved saddle-point equations at finite $d$, we find that
it does not, at the level of the annealed free energy: the average is
controlled for $d\gg1$, but at $d=2$ it enforces a flavor symmetry that the
saddle breaks. At large
$\kappa$ the lowest-action saddle among those we examine spontaneously breaks
the flavor-permutation symmetry: one integrable flavor carries the
quasi-integrable phase and the remaining $d-1$ stay empty. Up to
power-law-small spectator corrections this saddle is the $d=1$ BBJM
saddle with a rescaled coupling, so the transition survives at
$\kappa_c(d)\simeq\sqrt d\,\kappa_c(1)$, to within a few percent. It leaves
the physical window $\kappa\le1$ only for $d>d_c$; at $\beta J=110$,
$d_c\simeq33.5$, scaling as $0.30\,\beta J$ away from the critical point, and
$d_c\simeq4.9$ at the critical temperature. We map the
phase diagram, verify the $\sqrt d$ law for $d=1,2,3,5,10$ (full system up to
$d=5$), and show that the chord number grows quadratically in the broken phase and
linearly in the chaotic phase, while the Lanczos coefficients change from
linear to square-root growth at the transition. The
color-averaged model is well defined as a chord model, but as a free-energy
ansatz it enforces flavor symmetry and therefore predicts melting already at
$d=2$, whereas the color-resolved model melts only at
$d\simeq d_c=1/\kappa_c(1)^2$.
\end{abstract}

\maketitle

\section{Introduction}

The Sachdev--Ye--Kitaev (SYK) model
\cite{Kitaev2015,Sachdev:1992fk,Maldacena:2016hyu,Maldacena:2016upp,Chowdhury:2021qpy}
is one of the simplest tractable examples of a maximally chaotic quantum
system, and in its double-scaled limit (DSSYK) it becomes exactly solvable
through chord diagrams \cite{Berkooz:2018jqr,Berkooz:2018qkz,Berkooz2025,Lin:2022rbf}.
The chord description is not restricted to the SYK model itself: other
microscopic constructions, such as the Parisi hypercube model in the large-$d$
limit \cite{Berkooz:2023cqc,Jia:2024tii} and its Fock-space generalizations
\cite{Dai:2026ifl}, share the same chord algebra and the same spectral
properties. The chord formalism has by now become the standard tool for
studying deformations of DSSYK, spectral statistics
\cite{Polchinski:2016xgd,Saad:2018bqo}, multi-species chords
\cite{Berkooz:2024ifu}, and operator growth
\cite{Cotler:2016fpe,Parker:2018yvk}.

A particularly interesting deformation is the interpolation between DSSYK and
an integrable commuting-SYK Hamiltonian proposed by Berkooz, Brukner, Jia and
Mamroud (BBJM) \cite{Berkooz:2024evs,Berkooz:2024ofm},
\begin{equation}
  \hat H_{\rm BBJM} = \nu\,\hat H_{\rm DSSYK} + \kappa\,\hat H_{\rm int},
  \qquad \nu^2+\kappa^2=1 .
\end{equation}
In the double-scaling limit the model is described by two species of chords,
``chaotic'' ($n$) and ``integrable'' ($z$), with crossing weights
$q_{nn}=q_{nz}=q=e^{-\lambda}$ and $q_{zz}=1$. At low temperature the free
energy has two branches, one continuously connected to the chaotic model at
$\kappa=0$ and one to the integrable model at $\kappa=1$, and the two branches
exchange dominance through a first-order transition \cite{Berkooz:2024evs}.
The transition line in the $(\kappa,1/\beta J)$ plane ends at a critical
point; at low temperature it follows $\kappa_c\simeq 2/\sqrt{\beta J}$
\cite{Berkooz:2024evs}. The dynamical signatures of this transition have been
studied in \cite{Aguilar-Gutierrez:2026jjv}: the chord number, which is
proportional to the Krylov state complexity in the classical limit, grows
linearly in time in the chaotic phase and quadratically in the
quasi-integrable phase, while the Krylov operator complexity grows
exponentially and quadratically, respectively.

A natural way to build chaos out of integrable blocks was proposed by Gao,
Lin and Peng (GLM) \cite{Gao:2024lve}. Their $d$-commuting SYK model combines
$d$ copies of the commuting-SYK Hamiltonian,
\begin{equation}
  \hat H_{\rm dcSYK} = \frac{1}{\sqrt d}\sum_{a=1}^{d}\hat H_a ,
\end{equation}
where each $\hat H_a$ is integrable but different copies do not commute. In
the double-scaling limit the multi-color chord rules are simple: two chords of
the same color cross with weight $1$, while two chords of different colors
cross with weight $q$. Averaging over the $d$ colors gives the effective
weight
\begin{equation}
  \bar q(d) = \frac1d + \left(1-\frac1d\right)q ,
  \label{eq:qbar}
\end{equation}
which interpolates between the integrable value $\bar q(1)=1$ and the chaotic
value $\bar q(\infty)=q$. GLM studied this model in two limits. For $q=0$ the
spectrum has a square-root edge above a crossover temperature
$T_{\rm GLM}=J/(2d)$ and an exponential tail below it, and they emphasize that
this is a spectral crossover, not a thermodynamic transition. For $q\to1$ and
$d<9$ they argue instead that the crossover temperature has a thermodynamic
meaning: self-averaging holds only for $T\gtrsim T_c$, and a new phase appears
below $T_c$ (see Ref.~\cite{Gao:2024lve}). Our chord description is the $\lambda\to0$ semiclassical
limit of the chord path integral, i.e.\ the $q\to1$ regime, and its
saddle-point equations are $\lambda$-independent at leading order; the
comparison below is therefore with GLM's $q\to1$ case. As $d\to\infty$ the
intersection matrix becomes uniform and the model approaches DSSYK at the
level of the chord rules; the spectrum of the integrable sector itself is
Gaussian by the central limit theorem applied to the $d$ copies.
Crucially, \eqref{eq:qbar} was derived in \cite{Gao:2024lve} as the leading
large-$d$ correction to the moments, under the assumption $d\gg n$ for an
$n$-chord diagram; it is not an exact statement about the saddle-point free
energy at $d=2$. The melting conclusion is therefore an inference from
combining the averaged weight of \cite{Gao:2024lve} with the two-chord model
of \cite{Berkooz:2024ofm}; neither reference states it for the multi-flavor
model. Their construction is what makes the question sharp: it provides a
controlled family with a non-commuting integrable sector, and $\bar q(d)$ is a
natural large-$d$ benchmark.

The question we address in this paper is the following. What happens to the
BBJM first-order transition if the single integrable flavor is replaced by the
$d$ non-commuting integrable flavors of the $d$-commuting construction? A
first guess, based on \eqref{eq:qbar}, is that the integrable sector becomes
progressively more chaotic and that the transition is eventually destroyed.
Indeed, in the two-chord language of \cite{Berkooz:2024ofm} the replacement
$q_{zz}\to\bar q(d)<1$ corresponds to turning on the symmetric deformation
$l_{zz}=-\ln\bar q(d)/\lambda$, and \cite{Berkooz:2024ofm} showed that a large
enough $l_{zz}$ removes the transition altogether. Since $\bar q(2)=0.684$ for
$q=e^{-1}$, one might conclude that a single extra flavor already melts the
transition. We have checked this directly: solving the two-field model with
$l_{nz}=1$ and $l_{zz}=-\ln\bar q(2)/\lambda\simeq0.38$ gives a single
smooth branch, with no first-order transition.

In this paper we show that this conclusion is a property of the color-averaged
model when extrapolated to small $d$, rather than of the microscopic
multi-color model: it follows from
enforcing flavor symmetry before extremizing. The color-resolved chord path integral admits a saddle that
spontaneously breaks the permutation symmetry of the $d$ integrable flavors:
one flavor carries the large chord number of the quasi-integrable phase, while
the other $d-1$ flavors remain at zero. Because the spectators do not
contribute, this saddle reduces to the $d=1$ BBJM saddle with the replacement
$\kappa\to\kappa/\sqrt d$, up to power-law-small corrections. Consequently the transition survives, but moves to
larger $\kappa$:
\begin{equation}
  \kappa_c(d,\beta J) = \sqrt d\;\kappa_c(1,\beta J) .
  \label{eq:sqrtd}
\end{equation}
It disappears from the physical window $\kappa\le1$ only when
\begin{equation}
  d > d_c(\beta J) = \frac{1}{\kappa_c(1,\beta J)^2} .
  \label{eq:dc}
\end{equation}
At $\beta J=110$ we find $\kappa_c(1)=0.173$ and $d_c\simeq 33.5$; the
numerical data over $30\le\beta J\le240$ are well reproduced by
$\kappa_c(1)\simeq 1.873/\sqrt{\beta J}-0.634/\beta J$ (two-parameter
least-squares fit over $30\le\beta J\le240$, residuals below $0.3\%$), so
that $d_c\simeq0.30\,\beta J$. The transition is therefore not melted at $d=2$; it
survives up to $d\sim\mathcal O(10^1)$ at low temperature, and for any fixed
$d$ it reappears upon lowering the temperature. Because $d_c$ is bounded below
by its value at the critical point, $d_c(T_*)\simeq4.9$, at least five flavors
are needed for the transition to be removable at all.

The two-chord deformation studied in \cite{Berkooz:2024ofm} and the
color-resolved model considered here are genuinely different: the former
imposes $g_{z_1}=\dots=g_{z_d}$ at the level of the saddle, while the latter
allows the saddle to choose. We verify numerically, for $d=2,3,5$ with the full system and for $d=10$
by continuation of the symmetric branch, that the flavor-symmetric saddle
never has the lowest on-shell action among the saddles we examined for
$\kappa>\kappa_c(d)$,
and that the broken saddle reproduces the $d=1$ result to better than
$10^{-3}$ in $S$ at $d=2$ and $d=5$. We also show that the dynamical
state-complexity signature of the transition survives: the chord number grows
as $t^2$ in the
broken phase and linearly in the chaotic phase, with the same universal jump
at $\kappa_c(d)$ for every $d$.

The paper is organized as follows. In Sec.~\ref{sec:model} we define the model
and its exact chord rules, and explain why the color-averaged weight
\eqref{eq:qbar} does not determine the free energy. In Sec.~\ref{sec:saddles}
we derive the multi-field action and its on-shell value, and construct the
chaotic, symmetric and flavor-broken saddles. In Sec.~\ref{sec:thermo} we
obtain the phase diagram, the $\sqrt d$ law \eqref{eq:sqrtd} and the melting
line \eqref{eq:dc}. In Sec.~\ref{sec:dynamics} we study the real-time
evolution of the chord number across the transition. We conclude in
Sec.~\ref{sec:conclusions}. Appendix~\ref{app:action} contains the derivation
of the multi-field action and of its on-shell value, and
Appendix~\ref{app:numerics} the numerical methods and checks.

\section{The model and its chord description}
\label{sec:model}

\subsection{Single-flavor BBJM model}

We begin by recalling the BBJM model \cite{Berkooz:2024evs,Berkooz:2024ofm}.
The Hamiltonian is
\begin{equation}
  \hat H = \nu\,\hat H_{\rm DSSYK} + \kappa\,\hat H_{\rm int},
  \qquad \nu^2+\kappa^2=1 ,
  \label{eq:BBJM_H}
\end{equation}
where $\hat H_{\rm DSSYK}$ is the double-scaled SYK Hamiltonian with
$p$-body interactions and $\hat H_{\rm int}$ is an integrable Hamiltonian
built out of mutually commuting fermion bilinears \cite{Gao:2023gta}. In the double-scaling limit
\begin{equation}
  N,p\to\infty , \qquad \lambda \equiv \frac{2p^2}{N}\ \text{fixed} ,
\end{equation}
the ensemble-averaged moments of $\hat H$ are computed by chord diagrams with
two species of chords, chaotic ($n$) and integrable ($z$), with crossing
weights
\begin{equation}
  q_{nn}=q_{nz}=q=e^{-\lambda} , \qquad q_{zz}=1 .
  \label{eq:BBJM_weights}
\end{equation}

The free energy is determined by the saddle point of the chord path integral.
In the classical limit $\lambda\to0$ the saddle is described by two bilocal
fields $g_n,g_z$, which depend only on $\tau=\tau_1-\tau_2$ at the saddle, and
obey the Liouville equations
\begin{equation}
  \partial_\tau^2 g_n = 2J^2\nu^2\, e^{g_n+g_z} ,
  \qquad
  \partial_\tau^2 g_z = 2J^2\kappa^2\, e^{g_n} ,
  \label{eq:BBJM_eom}
\end{equation}
with $g_n(0)=g_n(\beta)=g_z(0)=g_z(\beta)=0$. The on-shell action is
\begin{equation}
  S_{\rm on\text{-}shell}
  = \frac{\beta}{4}\int_0^\beta d\tau\,
  \Big[\nu^2(g_n-2)\,e^{g_n+g_z} + 2\kappa^2(g_n-1)\,e^{g_n}\Big] ,
  \label{eq:BBJM_onshell}
\end{equation}
and the total chord number is $\ell(\tau)=-(g_n+g_z)$. Equations
\eqref{eq:BBJM_eom}--\eqref{eq:BBJM_onshell} are the $\lambda_{zz}=0$ case of
the general two-chord formulas of \cite{Berkooz:2024ofm}, to which we will
return below.

\subsection{Multi-flavor model and exact chord rules}
\label{sec:multicolor}

We now replace the single integrable Hamiltonian by the $d$-commuting
construction \cite{Gao:2024lve},
\begin{equation}
  \hat H = \nu\,\hat H_{\rm DSSYK}
  + \frac{\kappa}{\sqrt d}\sum_{a=1}^{d}\hat H_a ,
  \qquad \nu^2+\kappa^2=1 .
  \label{eq:H}
\end{equation}
Each $\hat H_a$ is a commuting-SYK Hamiltonian; different copies do not
commute because their fermion bilinears overlap. The model has a
$\mathbb{Z}_d$ (more precisely $S_d$) permutation symmetry of the $d$
integrable flavors, together with the $\mathbb{Z}_2$ symmetry
$\kappa\to-\kappa$.

The chord description of \eqref{eq:H} follows from the rules of
\cite{Gao:2024lve}. There are $d+1$ species of chords: one chaotic flavor
($a=0$) and $d$ integrable flavors ($a=1,\dots,d$). Two chords of the same
integrable flavor cross with weight $1$; any other pair crosses with weight
$q=e^{-\lambda}$. Equivalently, the intersection matrix
$\M_{ab}=\lambda_{ab}/\lambda$ is
\begin{equation}
  \M_{00}=1 , \qquad \M_{0a}=\M_{a0}=1 , \qquad
  \M_{aa}=0 , \qquad \M_{ab}=1 \quad (a\ne b,\ a,b\ge1) .
  \label{eq:Mmatrix}
\end{equation}
The weights of \eqref{eq:H} enter through the couplings
\begin{equation}
  \kappa_0^2 = \nu^2 , \qquad \kappa_a^2 = \frac{\kappa^2}{d} \quad (a\ge1) .
  \label{eq:kappa_a}
\end{equation}
For $d=1$ this reduces to \eqref{eq:BBJM_weights}; for $d\to\infty$ the
intersection matrix becomes uniform, $\M_{ab}\to1$ for all $a\ne b$, and, as
far as their mutual crossings are concerned, the integrable flavors become
indistinguishable from the chaotic one.

\begin{figure}[t]
  \centering
  \includegraphics[width=0.92\textwidth]{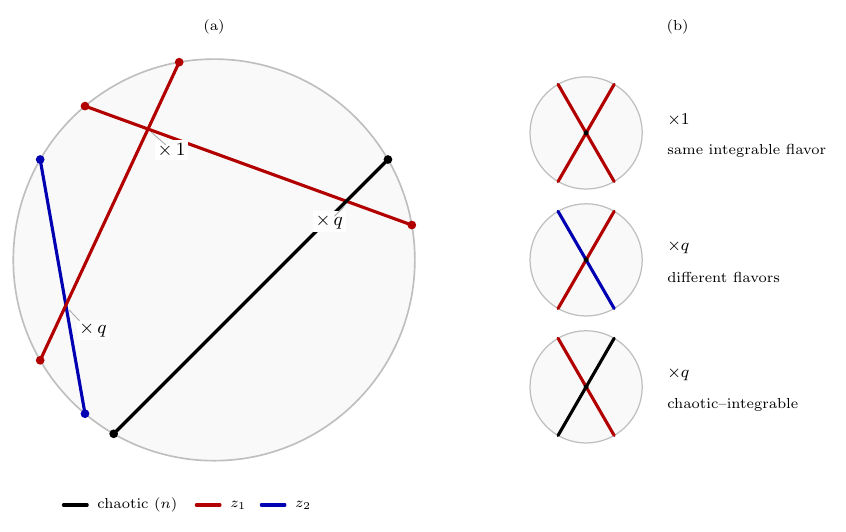}
  \caption{(a) Color-resolved chord diagram. Black chords carry the chaotic
  flavor $n$; colored chords carry integrable flavors $z_1$ and $z_2$ (the
  two red chords are two chords of the same flavor $z_1$). (b) The three
  crossing rules: two chords of the same integrable flavor carry weight $1$,
  while crossings between different integrable flavors and between the
  chaotic and an integrable flavor carry weight $q=e^{-\lambda}$. Summing
  over colorings is not the same as assigning the average weight
  \eqref{eq:qbar} to every $z$--$z$ crossing, because the colors are
  correlated along each chord.}
  \label{fig:chord}
\end{figure}

Since the entries of \eqref{eq:Mmatrix} are $0$ or $1$, the parameter
$\lambda$ drops out of the saddle-point equations and of the phase diagram at
leading order. We work in the $\lambda\to0$ semiclassical limit of the chord
path integral \cite{Berkooz:2024ofm}, in which the saddle-point approximation
is controlled by the $1/\lambda$ in the exponent of \eqref{eq:Z}; the
corrections to the saddle are $\mathcal O(\lambda)$. In the numerical evaluation we set $J=1$
and quote temperatures through $\beta J$;
the value $q=e^{-1}$ is used only when comparing with the color-averaged
weight \eqref{eq:qbar}. The resulting color-resolved chord structure is
illustrated in Fig.~\ref{fig:chord}: two chords of the same integrable flavor
cross with weight $1$, while every other crossing carries weight $q$.

\subsection{Why the color-averaged weight is not the free energy}
\label{sec:qbar}

It is tempting to replace, in the two-chord description, every $z$--$z$
crossing by the average weight \eqref{eq:qbar}, i.e.\ to set
$q_{zz}\to\bar q(d)$ and to work with the two-field action of
\cite{Berkooz:2024ofm}. This is what we will call the \emph{color-averaged}
model. It is a well-defined chord model, and for $d\gg1$ it captures the
leading correction to the moments \cite{Gao:2024lve}. It is, however, not
equivalent to the color-resolved model of Sec.~\ref{sec:multicolor}, for two
reasons.

First, \eqref{eq:qbar} is an average over crossings, not an average over
diagrams. For a fixed diagram with $C_{zz}$ crossings among integrable chords,
the sum over colorings gives a weight $q^{\,C_{zz}-C_{\rm same}}$, where
$C_{\rm same}$ is the number of crossings between chords of the same color.
The replacement $q^{\,C_{zz}}\to\bar q^{\,C_{zz}}$ is exact only when the
colors of crossing chords are independent, which requires $d\gg n$. At $d=2$
the two notions differ substantially: the exact weight between two different
colors is $q=e^{-1}=0.368$, while $\bar q(2)=0.684$.

Second, and more importantly, the saddle-point free energy is not linear in
the crossing weights. The color-resolved path integral is a sum over
field configurations, and the lowest-action configuration may break the color
symmetry. Enforcing $g_{z_1}=\dots=g_{z_d}$ before extremizing restricts the
variational space and can change the answer qualitatively. This is exactly
what happens here, as we show in Sec.~\ref{sec:saddles}. The color-averaged model is not an approximation we introduce to criticize:
it is the correct leading large-$d$ description of the moments, and it is
expected to be accurate wherever the flavor-broken saddle is subdominant. The
window in which the two saddles compete is precisely where the average is not
controlled, and it is this window that determines the phase diagram.

\section{Saddle points of the chord path integral}
\label{sec:saddles}

\subsection{Multi-field action and on-shell formula}
\label{sec:action}

The partition function is
\begin{equation}
  Z(\beta) = \int \prod_{a=0}^{d}\mathcal{D} n_a\,
  \exp\left[-\frac1\lambda S[n]\right] ,
  \label{eq:Z}
\end{equation}
with the same measure and normalization as in the two-chord case
\cite{Berkooz:2024ofm}. The chord expansion computes the average of the moments
$\langle\Tr\hat H^{k}\rangle$, i.e.\ the \emph{annealed} free energy
$-\ln\langle Z\rangle$. The quenched free energy $-\langle\ln Z\rangle$
differs by terms that are subleading in the double-scaling limit, where the
free energy is self-averaging at large $N$; we work in the standard annealed
setting of \cite{Berkooz:2024ofm,Aguilar-Gutierrez:2026jjv}. The flavor-broken
saddle has $d$ degenerate vacua; since $d$ is held fixed as $N\to\infty$, the
associated $\ln d$ entropy is subleading in the double-scaling limit and does
not affect the leading free energy. More generally, the annealed and quenched
free energies differ by $\mathcal O(1/N)$ (of which the $(\ln d)/N$ term is
the leading piece), so the phase diagram below is a leading-order statement;
a quenched analysis would be needed to locate the transition at subleading
order. A heuristic estimate of the induced shift is
$\delta\kappa_c/\kappa_c\sim\mathcal O(1/(N\ell))$, since the free-energy
difference is $\mathcal O(1/N)$ and the transition is first order with a
latent heat of order $\ell$; this is negligible at large $N$ and does not
affect the leading phase diagram. With the chord action
\begin{multline}
  S[n] = \frac14\int_0^\beta d\tau_1\int_0^\beta d\tau_2
  \int_{\tau_1}^{\tau_2}d\tau_3\int_{\tau_2}^{\tau_1}d\tau_4\,
  \sum_{a,b=0}^{d}\M_{ab}\, n_a(\tau_1,\tau_2)\,n_b(\tau_3,\tau_4) \\
  + \frac12\int_0^\beta d\tau_1\int_0^\beta d\tau_2\,
  \sum_{a=0}^{d} n_a(\tau_1,\tau_2)
  \left[\ln\frac{n_a(\tau_1,\tau_2)}{\kappa_a^2J^2}-1\right] .
  \label{eq:S}
\end{multline}

Here $n_a(\tau_1,\tau_2)$ is the density of chords of species $a$ connecting
$\tau_1$ and $\tau_2$. We have set $\Tr\mathbb{1}=1$ and keep the coupling $J$ explicit, setting
$J=1$ only when quoting numbers; we use
$q=e^{-\lambda}$ and $\M_{ab}=\lambda_{ab}/\lambda$ throughout. Introducing
\begin{equation}
  g_a(\tau_1,\tau_2) = -\int_{\tau_1}^{\tau_2}d\tau_3
  \int_{\tau_2}^{\tau_1}d\tau_4\, n_a(\tau_3,\tau_4) ,
  \label{eq:g_def}
\end{equation}
so that $n_a=-\frac12\partial_1\partial_2 g_a$, and varying \eqref{eq:S}
gives the saddle-point equations
\begin{equation}
  \partial_1\partial_2\, g_a(\tau_1,\tau_2)
  = -2J^2\kappa_a^2\,
  \exp\left[\sum_{b=0}^{d}\M_{ab}\, g_b(\tau_1,\tau_2)\right] .
  \label{eq:eom}
\end{equation}
At the saddle the fields depend only on $\tau=\tau_1-\tau_2$, and
\eqref{eq:eom} becomes
\begin{equation}
  \partial_\tau^2 g_a = 2J^2\kappa_a^2\,
  \exp\left[\sum_{b=0}^{d}\M_{ab}\, g_b\right] ,
  \qquad g_a(0)=g_a(\beta)=0 .
  \label{eq:eom_tau}
\end{equation}
The total chord number is $\ell=-\sum_{a=0}^{d}g_a$, which we quote at the
midpoint $\tau=\beta/2$ in all the tables and figures below.

Substituting the solution back into \eqref{eq:S} gives the on-shell action
\begin{equation}
  S_{\rm on\text{-}shell} = \frac{\beta}{8}\int_0^\beta d\tau\,
  \sum_{a,b=0}^{d}\M_{ab}\, g_a''(\tau)\,g_b(\tau)
  - \frac{\beta}{2}\int_0^\beta d\tau\,
  \sum_{a=0}^{d}\kappa_a^2J^2\,
  \exp\left[\sum_{b=0}^{d}\M_{ab}\,g_b(\tau)\right] .
  \label{eq:onshell}
\end{equation}
For $d=1$, \eqref{eq:onshell} reduces to \eqref{eq:BBJM_onshell}; the
derivation is given in Appendix~\ref{app:action}, where we also show that the
two forms differ only by a total derivative that vanishes on shell.

\subsection{The chaotic saddle}

The first saddle is the one continuously connected to the DSSYK solution at
$\kappa=0$. For small $\kappa$ we can expand
$g_n=g_n^{(0)}+\kappa^2g_n^{(1)}+\dots$ and $g_a=(\kappa^2/d) g_a^{(1)}+\dots$,
with \cite{Berkooz:2024ofm}
\begin{equation}
  g_n^{(0)}(\tau) = 2\ln\left[
  \frac{\cos\frac{\pi v}{2}}{\cos\left(\pi v\left(\frac12-\frac{\tau}{\beta}\right)\right)}
  \right] , \qquad
  \beta J = \frac{\pi v}{\cos\frac{\pi v}{2}} ,
  \label{eq:chaotic_g}
\end{equation}
and $g_a^{(1)}=g_n^{(0)}$ for every integrable flavor, since at order
$\kappa^2$ the linearized equation for $g_a$ is
$(g_a^{(1)})''=2J^2e^{g_n^{(0)}}$, which is precisely the $\kappa=0$ equation
satisfied by $g_n^{(0)}$ itself. The total chord number
is $\ell(\beta/2)=-g_n(\beta/2)\simeq7.1$ at $\beta J=110$, and the on-shell
action is
\begin{equation}
  S_{\rm ch} = -2\beta J + \frac{\pi^2}{2} + \mathcal{O}\left((\beta J)^{-1}\right)
  \simeq -215.1 \quad (\beta J=110) .
  \label{eq:Sch}
\end{equation}
This branch can be followed to $\kappa=1$; its chord number stays at
$\ell\sim7$--$13$ for $d=2$--$10$, and it is the branch that the double-field
calculation of the color-symmetric ansatz follows.

\subsection{The flavor-broken saddle}
\label{sec:broken}

The second saddle is the one that carries the quasi-integrable phase. Consider
the ansatz in which a single integrable flavor, say $a=1$, is non-trivial
while $g_a=0$ for $a=2,\dots,d$. The equations \eqref{eq:eom_tau} become
\begin{equation}
  \partial_\tau^2 g_n = 2J^2\nu^2\, e^{g_n+g_1} ,
  \qquad
  \partial_\tau^2 g_1 = 2J^2\frac{\kappa^2}{d}\, e^{g_n} ,
  \qquad
  \partial_\tau^2 g_a = 2J^2\frac{\kappa^2}{d}\, e^{g_n+g_1} \quad (a\ge2) .
  \label{eq:broken_eom}
\end{equation}
The spectator equations are solved by $g_a=0$ up to corrections of order
$e^{g_n+g_1}$. These are exponentially small at the midpoint of the thermal
circle, but they are largest in a boundary layer near $\tau=0$ and $\tau=\beta$
where $g_1$ has not yet grown, so that the correction to the on-shell action
is power-law-small rather than exponential; we quantify it below.
The first two equations are exactly the BBJM equations \eqref{eq:BBJM_eom}
with the replacement
\begin{equation}
  \kappa^2 \;\longrightarrow\; \kappa_{\rm eff}^2 \equiv \frac{\kappa^2}{d} .
  \label{eq:kappa_eff}
\end{equation}
Therefore, up to power-law-small spectator corrections, the broken
saddle is the $d=1$ BBJM saddle at $\kappa_{\rm eff}=\kappa/\sqrt d$:
\begin{equation}
  S_{\rm br}(\kappa;\beta J) \simeq S_{\rm BBJM}\left(\frac{\kappa}{\sqrt d};\beta J\right) ,
  \qquad
  \ell_{\rm br}(\kappa;\beta J) \simeq \ell_{\rm BBJM}\left(\frac{\kappa}{\sqrt d};\beta J\right) .
  \label{eq:mapping}
\end{equation}
The right-hand side is the \emph{normalized} BBJM saddle at
$\kappa_{\rm eff}=\kappa/\sqrt d$, i.e.\ with
$\nu_{\rm eff}=\sqrt{1-\kappa^2/d}$, whereas the broken saddle of the
original model has $\nu=\sqrt{1-\kappa^2}$; the difference enters only
through the source of the frozen chaotic flavor and is power-law-small,
$\mathcal O\!\left((\beta J)^{-2}\right)$.
At $\kappa=1$ and $d=2$ this gives $\ell_{\rm br}=1512$ and
$S_{\rm br}=-3026$, to be compared with $\ell_{\rm ch}=12.9$ and
$S=-302$ for the color-symmetric saddle (the $\kappa=1$ limit of the chaotic
branch). At $\kappa=1$ one has $\nu=0$
and hence $g_n=0$; the symmetric saddle then has two equal integrable fields
with $g_z(\beta/2)\simeq-6.47$, so that $\ell=2\times6.47=12.9$. The broken saddle is lower by a
factor of ten in the free energy.

\begin{figure}[t]
  \centering
  \includegraphics[width=0.95\textwidth]{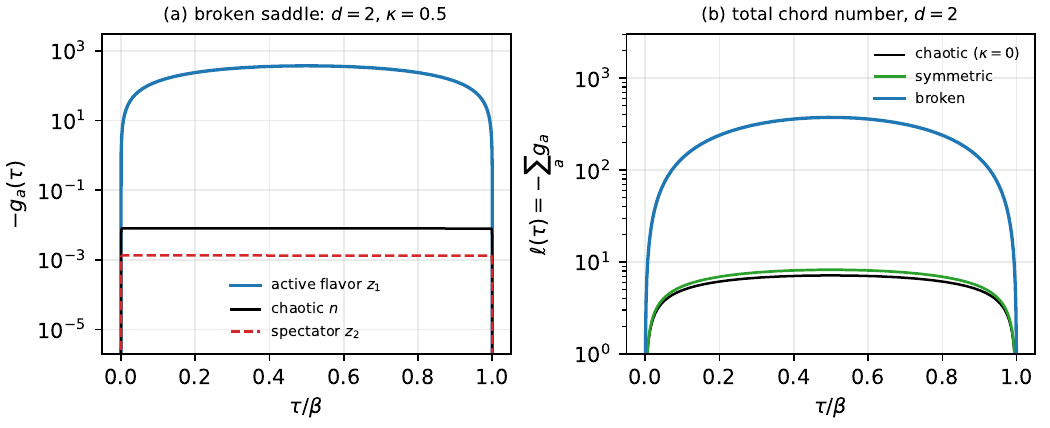}
  \caption{(a) The fields of the flavor-broken saddle at $d=2$, $\kappa=0.5$,
  $\beta J=110$: the active integrable flavor carries $-g_{z_1}\simeq375$ at
  the midpoint, while the chaotic flavor ($-g_n\simeq8\times10^{-3}$) and the
  spectator ($-g_{z_2}\simeq1.3\times10^{-3}$) stay essentially empty.
  (b) The total chord number $\ell(\tau)$ for the chaotic saddle ($\kappa=0$,
  $\ell=7.1$), the symmetric saddle ($d=2$, $\kappa=0.5$, $\ell=8.2$) and the
  broken saddle ($d=2$, $\kappa=0.5$, $\ell=374.6$). The symmetric ansatz is
  the continuation of the chaotic branch; only the broken saddle develops a
  large chord number.}
  \label{fig:profiles}
\end{figure}

The physical interpretation is spontaneous breaking of the flavor-permutation
symmetry in the large-$N$ saddle-point sense: the ensemble is color symmetric,
but at large $\kappa$ the saddle
chooses one flavor to carry the quasi-integrable phase. The resulting field
profiles are shown in Fig.~\ref{fig:profiles}(a). The $d-1$ spectators
are not simply inactive by fiat; their equations force them to vanish because
the exponent $g_n+g_1$ is large and negative. This is the chord-space analogue
of a vector-model condensate. The same observation explains why no saddle with two large flavors exists.
Once one flavor has grown, $g_1\ll0$, the source of every other integrable
flavor is proportional to $e^{g_n+g_1}$ and is exponentially suppressed; a
second flavor therefore cannot grow, and the $k$-active saddles with $k\ge2$
collapse back to the small-chord solution (Table~\ref{tab:kactive}). In this
sense the one-flavor-broken saddle is the only large-$\ell$ configuration
available to the model, which is the qualitative reason behind its dominance.

\subsection{The color-symmetric saddle}

For completeness, consider the symmetric ansatz
$g_1=\dots=g_d\equiv g_z$. Then
\begin{equation}
  \partial_\tau^2 g_n = 2J^2\nu^2\, e^{g_n+d\,g_z} ,
  \qquad
  \partial_\tau^2 g_z = 2J^2\frac{\kappa^2}{d}(d-1)\,
  e^{g_n+(d-1)g_z} ,
  \label{eq:sym_eom}
\end{equation}
which is the color-averaged model with
$\bar q(d)$ replaced by the exact same-color weight; at large $d$ it reduces to
the $q_{zz}=q$ two-chord model. At $\kappa=0$ the ansatz gives $g_z=0$, i.e.\
the chaotic saddle, and it is the continuation of that branch to finite
$\kappa$; we therefore refer to it as the \emph{compact} branch. It exists for
all $\kappa$, but we have checked numerically, for $d=2,3,5$ with the full
system and for $d=10$ by continuation of the symmetric branch, that it never
has the lowest action for $\kappa>\kappa_c(d)$: at $\kappa=0.5$ and $d=2$ its
action is $-221$, while the broken saddle gives $-763$. The reason is simple:
the symmetric ansatz forces all $d$ flavors to share the chord number, which
costs a factor $d$ in the exponent, whereas the broken saddle lets one flavor
grow freely. The symmetric saddle is the chord-space counterpart of the
color-averaged model, and it is precisely the saddle that the analysis of
\eqref{eq:qbar} follows. Fig.~\ref{fig:profiles}(b) compares the total chord
number of the three saddles.

Collecting these results: among the uniform saddles examined here, only two
carry a large chord number, the compact branch above — which contains the
chaotic saddle and, as we show in Sec.~\ref{sec:stability}, every $k$-active
saddle with $k\ge2$ — and the flavor-broken branch of
Sec.~\ref{sec:broken}. The first-order transition is the crossing of these two
branches, and the competition between them is the whole content of
Sec.~\ref{sec:thermo}.

\subsection{Numerical verification}

We solved \eqref{eq:eom_tau} numerically with a Chebyshev--Newton method
(Appendix~\ref{app:numerics}), both for the full $(d+1)$-field system and for
the reduced two-field system of the broken ansatz. For $d=2$ at
$\beta J=110$ and $\kappa=0.5$ the full three-field solution and the $d=1$
saddle at $\kappa_{\rm eff}=0.5/\sqrt2=0.35355$ give the same on-shell action
$S\simeq-763.4$ and the same $\ell=374.6$; their difference is
$3\times10^{-4}$, stable under changes of the grid (the absolute
normalization carries a discretization uncertainty of about $0.1$; see
Appendix~\ref{app:numerics}). At $\kappa=1$ the agreement is better than
$0.1\%$ ($-3026$ against $-3025$), and for $d=5$ at $\kappa=0.8$ the two
actions again differ by only $3\times10^{-4}$ ($-781.4$).

The accuracy of this mapping deserves a comment, because the two models are
not identical term by term. The active pair obeys \eqref{eq:broken_eom} with
the \emph{same} $\nu=\sqrt{1-\kappa^2}$ as the original Hamiltonian, whereas
the normalized BBJM model at $\kappa_{\rm eff}=\kappa/\sqrt d$ would have
$\nu_{\rm eff}=\sqrt{1-\kappa^2/d}$; the two differ in the chaotic coupling.
In the broken phase, however, the chaotic flavor is frozen: the source in its
equation is $e^{g_n+g_z}$ with $g_z(\beta/2)\simeq-\ell$, so
$g_n$ never grows beyond the level it reaches inside a boundary layer of
width $\sim d/(\kappa^2\beta J)$ near $\tau=0$, and the on-shell action is
independent of $\nu$ up to that accuracy. Concretely, for $d=2$ and
$\kappa=0.5$ the difference
$S_{\rm br}-S_{\rm BBJM}(\kappa/\sqrt d)$ is $3.4\times10^{-4}$ at
$\beta J=110$, decreasing as $(\beta J)^{-2}$ ($6.4\times10^{-3}$ at
$\beta J=30$, $8.3\times10^{-5}$ at $\beta J=220$); at fixed
$\kappa/\sqrt d$ it grows linearly with $d$. The accuracy is therefore a
power law rather than an exponential, because the region where $e^{g_z}$ is
of order one is a boundary layer and not the bulk. It is nonetheless
parametrically smaller than the $\mathcal O(\lambda)$ corrections to the
saddle-point approximation, which is why the $d=1$ results can be used
directly.

\begin{table}[t]
  \centering
  \begin{tabular}{c|c|c|c|c|c}
    \hline
    $d$ & $\kappa_c(d)$ & $\ell_{\rm ch}(\kappa_c)$ & $\ell_{\rm br}(\kappa_c)$
        & $S_{\rm ch}(\kappa_c)$ & $S_{\rm br}(\kappa_c)$ \\
    \hline
    1  & 0.173 & 7.15 & 70.4 & $-215.6$ & $-215.5$ \\
    2  & 0.244 & 7.16 & 70.5 & $-215.7$ & $-215.6$ \\
    3  & 0.299 & 7.17 & 70.6 & $-215.8$ & $-215.8$ \\
    5  & 0.387 & 7.18 & 70.9 & $-216.0$ & $-216.0$ \\
    10 & 0.548 & 7.22 & 71.3 & $-216.5$ & $-216.5$ \\
    \hline
  \end{tabular}
  \caption{Transition points at $\beta J=110$, together with the chord numbers
  and on-shell actions of the two competing saddles at the transition. The
  jump $\ell_{\rm br}-\ell_{\rm ch}\simeq63$ is universal, while the
  transition point moves as $\sqrt d$. The crossing is located by linear
  interpolation of $\Delta S$ between neighboring grid points; the residual
  difference of the actions ($\lesssim0.1$) reflects this interpolation
  error, and $\kappa_c$ is uncertain at the $10^{-3}$ level.}
  \label{tab:transition}
\end{table}

\subsection{Selection of the dominant saddle}
\label{sec:stability}

The broken saddle is selected by comparing on-shell actions, so the first
question is whether it is stable against populating the other flavors. We have
performed two checks.

\paragraph{Global check: $k$-active saddles.} Consider a saddle in which $k$
of the $d$ integrable flavors share a common large profile $g$ while the
remaining $d-k$ stay at $g=0$. The equations are
\begin{equation}
  \partial_\tau^2 g_n = 2J^2\nu^2\,e^{g_n+k g} ,
  \qquad
  \partial_\tau^2 g = 2J^2\frac{\kappa^2}{d}\,
  e^{g_n+(k-1)g} ,
  \label{eq:kactive}
\end{equation}
and the on-shell action follows from \eqref{eq:onshell}. The results are
collected in Table~\ref{tab:kactive}. In every case the $k=1$ saddle has by
far the lowest action: at $d=2$, $\kappa=0.35$ it is lower by $170$ in $S$,
and at $d=3$, $\kappa=0.5$ by about $300$. Uniformly populating more flavors
is therefore strongly disfavored, which is the chord-space statement that the
flavor-permutation symmetry is spontaneously broken in the saddle-point sense.
For $k\ge2$ this is not merely a hierarchy but a no-go statement. Put
$L\equiv-(g_n+kg)>0$, so that $g_n=-L-kg$; the arguments of the exponentials
in \eqref{eq:kactive} are then $-(g_n+kg)=-L$ for the chaotic flavor and
$g_n+(k-1)g=-L-g$ for each of the $k$ active ones. Using $g\simeq-L/k$, the
latter is $\simeq-(k-1)L/k$, which vanishes for $k=1$ but is of order $L$ for
$k\ge2$.
Hence for $k\ge2$ the right-hand side of the second equation in
\eqref{eq:kactive} is exponentially small in $L$, whereas the left-hand side
$g''\sim L/\beta^2$ is only algebraically small; self-consistency then requires
$L=\mathcal O(\ln\beta^2)$, and no large-$\ell$ solution exists. The $k\ge2$
rows of Table~\ref{tab:kactive} ($\ell\simeq7.2$--$7.4$, $S\simeq-216$) are
therefore not a separate family of saddles: they are the chaotic saddle
itself. The $k=1$ action, by contrast, is that of a single integrable flavor,
$S_1\simeq-\beta^2\kappa^2/(2d)$, and it is lower than the chaotic value
$-2\beta J$ whenever $\beta J\kappa^2/(2d)\gtrsim2$; near $\kappa_c$ the two
approach each other and the comparison is numerical
(Table~\ref{tab:kactive}).
The comparison is restricted to the saddles we could construct. As a further
check, we have performed random-initial-condition searches at
$d=2$ ($\kappa=0.5$) and $d=3$ ($\kappa=0.35,0.5$), with $40$ random starts
each. About $40\%$ of the starts converge, the rest being discarded as
non-convergent; every converged solution is one of the $k$-active uniform
saddles, with actions between $-207$ and $-222$, none lower than the
corresponding broken saddle ($-763.6$ at $d=2,\kappa=0.5$; $-515.6$ at
$d=3,\kappa=0.5$; $-271.6$ at $d=3,\kappa=0.35$). We have repeated the search
at the transition points themselves, $\kappa_c(2)=0.244$ and
$\kappa_c(3)=0.300$: the converged solutions are the chaotic saddle
($S=-215.6$) and a higher saddle at $\ell\simeq16.5$ ($S=-199.6$), none below
the crossing value.

\begin{table}[t]
  \centering
  \begin{tabular}{c|c|c|c|c}
    \hline
    $d$ & $\kappa$ & $k$ & $S_k$ & $\ell_k(\beta/2)$ \\
    \hline
    2 & 0.35 & 1 & $-386.1$ & 177.9 \\
    2 & 0.35 & 2 & $-216.2$ & 7.2 \\
    \hline
    3 & 0.50 & 1 & $-515.3$ & 247.5 \\
    3 & 0.50 & 2 & $-217.4$ & 7.4 \\
    3 & 0.50 & 3 & $-218.2$ & 7.4 \\
    \hline
  \end{tabular}
  \caption{On-shell action of the uniform $k$-active-flavor saddles at $\beta J=110$.
  The one-flavor-broken saddle ($k=1$) has the lowest action among the saddles
examined. For $k\ge2$ no large-$\ell$ solution exists: the entries are the
chaotic saddle, to which the $k\ge2$ ansatz reduces.}
  \label{tab:kactive}
\end{table}

\paragraph{One-loop Hessian.} Let $f_a(\tau)=\delta n_a(\tau)$ be a
fluctuation of the chord density that depends only on the relative time. The
second variation of \eqref{eq:S} is
\begin{equation}
  \delta^2 S = \frac{\beta}{2}\int_0^\beta\!\! d\tau\!\int_0^\beta\!\! d\sigma\,
  \sum_{a,b}\M_{ab}\,f_a(\tau)f_b(\sigma)\,K(\tau,\sigma)
  + \frac{\beta}{2}\int_0^\beta d\tau\sum_a \frac{f_a(\tau)^2}{n_a(\tau)} ,
  \label{eq:hessian}
\end{equation}
with the crossing kernel $K(\tau,\sigma)=2\min(\tau,\sigma,\beta-\tau,\beta-\sigma)$.
Discretized on a grid of up to $240$ points, this quadratic form is
\emph{indefinite for every saddle}, the DSSYK saddle included: at
$\kappa=0.13$ the $d=1$ chaotic saddle has three negative modes, the
integrable one has none, and at $d=2$, $\kappa=0.35$ the broken saddle has
eight. The negative modes of the broken saddle are localized in the spectator
sector and arise from its mixing with the active pair through $\M_{ab}$; the
active block alone is positive definite, with the same lowest eigenvalue
($+30.2$) as the $d=1$ saddle at $\kappa_{\rm eff}$, and the most negative
direction stays inside the physical domain ($\delta n_a/n_a\ge-0.39$). The
broken saddle is therefore a genuine solution of the saddle-point equations
but not necessarily a local minimum --- as is the DSSYK saddle. The lesson is
not that the saddle is unstable, but that the Hessian is not the relevant
criterion here; we return to this in App.~\ref{app:action}.

A comment on the status of this selection is in order. For a complexified
action each saddle defines a Lefschetz thimble, and the leading contribution
is controlled by the on-shell action: the Morse index contributes a phase but
not the magnitude at leading order in $\lambda$. The DSSYK saddle is in the
same situation, and the standard practice in the chord literature
\cite{Berkooz:2024ofm,Aguilar-Gutierrez:2026jjv} is to select the dominant
saddle by its on-shell action; we follow it. The one-loop correction to the
transition point can be estimated: near $\kappa_c(d)$ the action difference
grows with slope $d(\Delta S)/d\kappa\simeq1.2\times10^{3}$ (for $d=2$ at
$\beta J=110$), while a one-loop correction to the free energy is
$\mathcal O(1)$ in these units, so the induced shift of $\kappa_c$ is of order
$10^{-3}$, well below the accuracy of the $\sqrt d$ law and of the phase
diagram. Throughout, ``dominant'' therefore means the lowest on-shell action
among the uniform saddles examined.

\section{Thermodynamics: the phase diagram}
\label{sec:thermo}

\subsection{The transition and the $\sqrt d$ law}

The dominant saddle at a given $\kappa$ is the one with the lower on-shell
action. Before turning to the numerical solution it is worth recording the
two-line estimate that makes the scaling law inevitable. In the deep broken
phase the broken saddle reduces to a single flavor with a Gaussian spectrum,
$S_{\rm br}\simeq-\beta^2\kappa^2/(2d)$ and
$\ell_{\rm br}\simeq\kappa^2\beta^2/(4d)$, both of which scale as $\beta^2$,
whereas the chaotic saddle has $S_{\rm ch}\simeq-2\beta J$, which scales as
$\beta$. Equating the two gives $\kappa_c=2\sqrt{d/\beta J}$ and hence
$d_c=\beta J/4$. The numerical values --- $\kappa_c(1)=0.173$ against
$2/\sqrt{\beta J}=0.191$, and $d_c=0.30\,\beta J$ against $0.25\,\beta J$ ---
differ from these estimates by the finite-temperature corrections of relative
order $10/\beta J$.
This is the physical content of the $\sqrt d$ law: at low temperature the
$\beta^2$ branch always wins eventually, and $\sqrt d$ shifts the crossing
point without removing it.

For $d=1$ the two branches \eqref{eq:Sch} and \eqref{eq:mapping}
cross at $\kappa_c(1,\beta J)$; at $\beta J=110$ we find
$\kappa_c(1)=0.173$, in rough agreement with the low-temperature estimate
$\kappa_c\simeq2/\sqrt{\beta J}=0.191$ of \cite{Berkooz:2024evs}, the
difference being the finite-temperature correction
$\kappa_c(1)\simeq(2/\sqrt{\beta J})(1-10/\beta J)$ estimated above. Because the
broken saddle is the $d=1$ saddle at $\kappa/\sqrt d$, the crossing condition
$S_{\rm ch}(\kappa)=S_{\rm br}(\kappa)$ is the same as the $d=1$ condition
with $\kappa\to\kappa/\sqrt d$. Using the fact that $S_{\rm ch}$ depends only
weakly on $\kappa$ (it changes by less than $0.5\%$ over the range of
$\kappa$ relevant here), we obtain
\begin{equation}
  \kappa_c(d,\beta J) \simeq \sqrt d\;\kappa_c(1,\beta J) ,
  \label{eq:kc}
\end{equation}

\begin{figure}[t]
  \centering
  \includegraphics[width=0.70\textwidth]{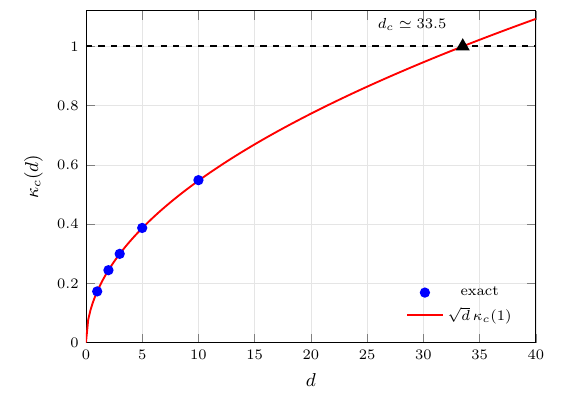}
  \caption{Transition point $\kappa_c(d)$ at $\beta J=110$. The dots are the
  color-resolved values (full system for $d=2,3,5$; mapping for $d=10$;
  Table~\ref{tab:transition}); the red line is
  $\sqrt d\,\kappa_c(1)$ with $\kappa_c(1)=0.173$. The dashed line is the
  boundary $\kappa=1$ of the physical parameter space, and the triangle marks
  $d_c\simeq33.5$, above which the transition lies outside $\kappa\le1$.}
  \label{fig:kcd}
\end{figure}

Table~\ref{tab:transition} and Fig.~\ref{fig:kcd} confirm this approximate
law for $d=1,2,3,5,10$ to better than $0.4\%$ at $\beta J=110$; the residual
is the sub-percent correction discussed below. We have checked it at
other temperatures as well: computing $\kappa_c(2)$ and $\kappa_c(3)$ directly
from the color-resolved model at $\beta J=30,60,160$, we find agreement with
$\sqrt d\,\kappa_c(1)$ at the level of $0.5\%$ ($d=2$) and $1.1\%$ ($d=3$) at
$\beta J=30$, and better than $0.3\%$ at $\beta J=60$ and $160$. The deviation
grows slowly with $d$ and with temperature; its sources are the
$\mathcal O(1/\beta J)$ corrections to $S_{\rm ch}$, the induced spectator
terms ($\mathcal O((\beta J)^{-2})$, Sec.~\ref{sec:broken}), and the
finite-$\kappa$ corrections to
the integrable branch. The $1/\beta J$ part scales roughly as $d/\beta J$,
while the residual $\sim0.3\%$ at $\beta J=60$ and $160$ is $d$-independent
and is dominated by the spectator and finite-$\kappa$ corrections. It does not affect the phase diagram at the accuracy of
Fig.~\ref{fig:phase}.

\subsection{The melting line}

The transition is visible in the physical parameter space only if
$\kappa_c(d,\beta J)\le1$, i.e.
\begin{equation}
  d \le d_c(\beta J) = \frac{1}{\kappa_c(1,\beta J)^2} .
  \label{eq:dc2}
\end{equation}
For $d>d_c$ the chaotic saddle has the lower action for every $\kappa\in[0,1]$,
and the transition is absent. This is the precise sense in which the
transition ``melts'' in the multi-flavor model: it is not erased at $d=2$; it
is pushed out of the window only at $d\sim d_c$.

\begin{figure}[t]
  \centering
  \includegraphics[width=0.70\textwidth]{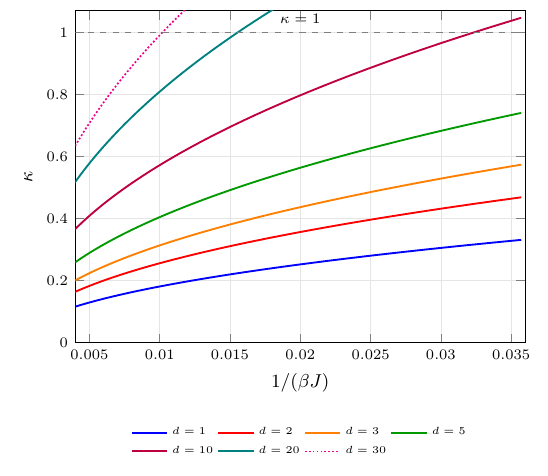}
  \caption{Phase diagram in the $(\kappa,1/\beta J)$ plane. Solid lines:
  transition points $\kappa_c(d,\beta J)$ for $d=1,2,3,5,10,20,30$, obtained
  from $\kappa_c(d)=\sqrt d\,\kappa_c(1)$ with the numerically determined
  $\kappa_c(1,\beta J)$. The broken (quasi-integrable) phase lies above each
  line. The dashed line is the boundary $\kappa=1$ of the physical parameter
  space.}
  \label{fig:phase}
\end{figure}

In Fig.~\ref{fig:phase} we show the phase diagram in the
$(\kappa,1/\beta J)$ plane for several values of $d$. Each line is
$\kappa_c(d,\beta J)$; the region above the line is the broken phase. The
horizontal dashed line is $\kappa=1$: the intersection of a given $d$-line
with it gives the temperature below which the transition is visible. For
$d=30$ this happens at $\beta J\simeq98$, while for $d=2$ the transition
remains inside the window down to the lowest temperature shown,
$\beta J\simeq28$.

The numerical values of $d_c$ are collected in Table~\ref{tab:dc}. The
dependence is approximately linear,
\begin{equation}
  d_c(\beta J) \simeq 0.30\,\beta J ,
  \label{eq:dcfit}
\end{equation}
the ratio $d_c/\beta J$ decreases slowly from $0.32$ at $\beta J=30$ to $0.30$
at $\beta J=240$. Equivalently,
for a fixed number of flavors the transition reappears below the temperature
$T_{\rm re}(d)\simeq0.30\,J/d$. This is the main thermodynamic result of this paper:
the multi-flavor transition is protected by flavor-symmetry breaking, and it
survives up to a number of flavors that grows linearly with $\beta J$.

\begin{table}[t]
  \centering
  \begin{tabular}{c|cccccccc}
    \hline
    $\beta J$ & 30 & 40 & 50 & 60 & 80 & 110 & 160 & 240 \\
    \hline
    $\kappa_c(1)$ & 0.320 & 0.280 & 0.252 & 0.231 & 0.202 & 0.173 & 0.144 & 0.118 \\
    $d_c$ & 9.7 & 12.7 & 15.7 & 18.7 & 24.6 & 33.5 & 48.3 & 72.1 \\
    \hline
  \end{tabular}
  \caption{The $d=1$ transition point and the melting line
  $d_c=1/\kappa_c(1)^2$ for several temperatures; $d_c$ is computed from the
  unrounded $\kappa_c(1)$.}
  \label{tab:dc}
\end{table}

\begin{figure}[t]
  \centering
  \includegraphics[width=0.60\textwidth]{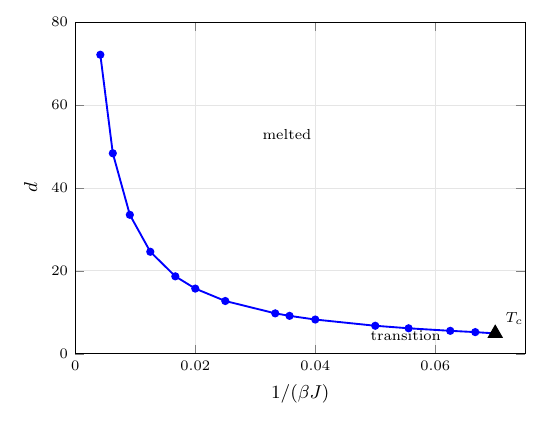}
  \caption{The melting line $d_c(\beta J)=1/\kappa_c(1,\beta J)^2$ in the
  $(d,1/\beta J)$ plane. Below the line the first-order transition is present
  at $\kappa_c(d)=\sqrt d\,\kappa_c(1)\le1$; above it the transition lies
  outside the physical window and the model is chaotic for all
  $\kappa\in[0,1]$. The triangle marks the critical point
  $(d_c,\beta J)=(4.9,13.5)$.}
  \label{fig:melting}
\end{figure}

The first-order line of the $d=1$ model does not extend to arbitrarily high
temperatures: it ends at a critical point. Following the two branches until
they merge, we locate it at
\begin{equation}
  (\kappa_*,\beta J_c) \simeq (0.45,\,13.5) ,
  \label{eq:critical}
\end{equation}
with no crossing for $\beta J\lesssim13.5$; the location is uncertain by
about $0.01$ in $\kappa_*$ and $0.5$ in $\beta J_c$, set by the resolution of
the branch merging. Because the broken saddle is the
$d=1$ saddle at $\kappa/\sqrt d$, the critical point of the $d$-line sits at
$(\sqrt d\,\kappa_*,\beta J_c)$: the critical temperature is independent of
$d$, while the critical value of $\kappa$ grows as $\sqrt d$. The melting
line therefore terminates at
\begin{equation}
  d_c(T_*) = \frac{1}{\kappa_*^2} \simeq 4.9\pm0.2 .
  \label{eq:dctc}
\end{equation}
Equivalently, at least five flavors are needed for the transition to be
removable at all: for $d\le4.9$ the line $\kappa_c(d)$ stays inside
$\kappa\le1$ at every temperature below $T_c$, so adding flavors pushes the
transition out of the window only when $d>4.9$.
This completes the phase diagram, which we show in the $(d,1/\beta J)$ plane
in Fig.~\ref{fig:melting}: for $d<d_c(T)$ the transition at $\kappa_c(d)$
lies inside $\kappa\le1$, while for $d>d_c(T)$ the model is chaotic for every
$\kappa\in[0,1]$. The melted region is bounded below by $d\simeq4.9$ at the
critical temperature and extends to $d\to\infty$ as $T\to0$.

\subsection{Universal jump and the role of the symmetric saddle}

\begin{figure}[t]
  \centering
  \includegraphics[width=0.70\textwidth]{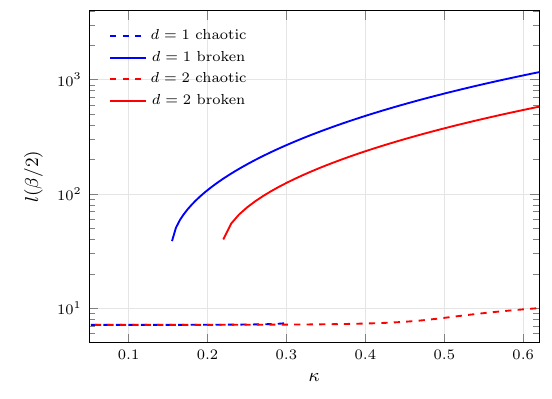}
  \caption{Chord number at the midpoint of the thermal circle as a function of
  $\kappa$ at $\beta J=110$. Dashed lines: chaotic branch. Solid lines:
  broken branch, which is the $d=1$ BBJM solution evaluated at
  $\kappa/\sqrt d$; it is drawn also below $\kappa_c$, where it is
  metastable. The transition is the crossing of the two branches, at
  $\kappa_c=0.173$ ($d=1$) and $0.244$ ($d=2$), where $\ell$ jumps by $63$.}
  \label{fig:lkappa}
\end{figure}

At the transition the chord number jumps from $\ell_{\rm ch}\simeq7$ to
$\ell_{\rm br}\simeq70$, independently of $d$ (Table~\ref{tab:transition}).
The jump is universal because the broken saddle is always the same $d=1$
solution evaluated at $\kappa_{\rm eff}=\kappa_c(1)$; only the value of
$\kappa$ at which this happens depends on $d$. In Fig.~\ref{fig:lkappa} we
plot $\ell(\beta/2)$ as a function of $\kappa$ for $d=1$ and $d=2$: the
chaotic branch stays at $\ell\simeq7$, while the broken branch, drawn also
below $\kappa_c$ where it is metastable, sits at $\ell\simeq70$ when it
crosses; the jump is about $63$.

It is worth emphasizing what happens to the color-symmetric saddle. At
$\kappa=1$ and $d=2$ it has $\ell=12.9$ and $S=-302$, while the broken saddle
has $\ell=1512$ and $S=-3026$. The symmetric saddle is therefore not merely
subdominant by a small amount: it is higher by an amount of order $\beta^2$,
which is the difference between a compact (DSSYK-like) and a Gaussian
(commuting-SYK-like) spectrum. This is the quantitative reason why the
color-averaged model, which follows the symmetric saddle, cannot capture the
correct phase diagram.

\subsection{Relation to the two-chord deformation of BBJM}
\label{sec:twoc}

The long paper of BBJM \cite{Berkooz:2024ofm} studied a two-chord model with
generic intersection weights -- the same class of deformations that underlies
the SYK renormalization-group flows of \cite{Anninos:2022qgy} -- and, in
particular, the deformation
$l_{zz}\neq0$ (in their notation $l_{zz}=\lambda_{zz}/\lambda$). They found
that for small $l_{zz}$ the transition survives and acquires a
zero-temperature endpoint, while for large $l_{zz}$ it disappears. The
color-averaged model of Sec.~\ref{sec:qbar} corresponds to
$l_{zz}=-\ln\bar q(d)/\lambda$, which grows with $d$ and reaches $1$ as
$d\to\infty$; in that model the transition is indeed removed at relatively
small $d$. Our result shows that this deformation is not the appropriate
effective description of the microscopic multi-color model: the $z$--$z$
crossings in \eqref{eq:Mmatrix}
are not a single symmetric deformation, because they occur only between
different colors. The dominant saddle avoids them by emptying all but one
flavor, and the effective description is the $d=1$ BBJM model with a rescaled
coupling. The two-chord deformation of \cite{Berkooz:2024ofm} is recovered
only if one insists on $g_{z_1}=\dots=g_{z_d}$, which is not what the path
integral does.

To connect this deformation to the operator-complexity diagnostic of
Sec.~\ref{sec:dynamics}, we followed the broken branch as a function of
$l_{zz}$. At $\kappa=0.5$ the two branches become degenerate at
$l_{zz}^*\simeq0.47$; below it the broken branch is the lowest among the
branches examined. As $l_{zz}\to1$ the collapse of $\tilde b_2$ disappears:
the coefficient rises continuously toward its chaotic value $\tilde b_2^2=2$
at $l_{zz}=1$, where the transition has disappeared. The exponential collapse
is therefore tied to the existence of the first-order transition, not to the
two-chord deformation alone. For $l_{zz}>0$ the broken saddle is not the
frozen-$n$ BBJM saddle, so the simple form \eqref{eq:bnbroken} is not expected
to hold quantitatively; we use only the qualitative statement that the
collapse tracks the transition.

\section{Dynamics}
\label{sec:dynamics}

\subsection{Real-time evolution of the chord number}

The state-complexity signature of the transition was studied for $d=1$ in
\cite{Aguilar-Gutierrez:2026jjv}; for background on Krylov complexity and its
use across integrable and chaotic systems see
\cite{Parker:2018yvk,Balasubramanian:2022tpr,Rabinovici:2022beu}. In the
classical limit the Krylov state complexity is proportional to the total chord
number,
\begin{equation}
  \mathcal C_K(t) \propto \ell(t)-\ell(0) , \qquad
  \ell(t) = -\sum_{a=0}^{d} g_a(t) ,
  \label{eq:CK}
\end{equation}
and its growth rate distinguishes the two phases. To compute $\ell(t)$ we
analytically continue the Euclidean saddle to real time,
$\tau\to\beta/2+it$, and solve
\begin{equation}
  \partial_t^2 g_a = -2J^2\kappa_a^2\,
  \exp\left[\sum_{b=0}^{d}\M_{ab}\,g_b\right] ,
  \qquad
  g_a(0)=g_a^{\rm E}\left(\frac\beta2\right) , \quad
  \dot g_a(0)=0 .
  \label{eq:realtime}
\end{equation}
We use the same numerical scheme as in \cite{Aguilar-Gutierrez:2026jjv}, with
an adaptive Runge--Kutta integrator. The continuation is the standard
semiclassical one of \cite{Aguilar-Gutierrez:2026jjv}: it evolves the saddle
configuration along the Lorentzian direction and provides the leading-order
real-time dynamics, with the same status as in that reference.

\begin{figure}[t]
  \centering
  \includegraphics[width=0.70\textwidth]{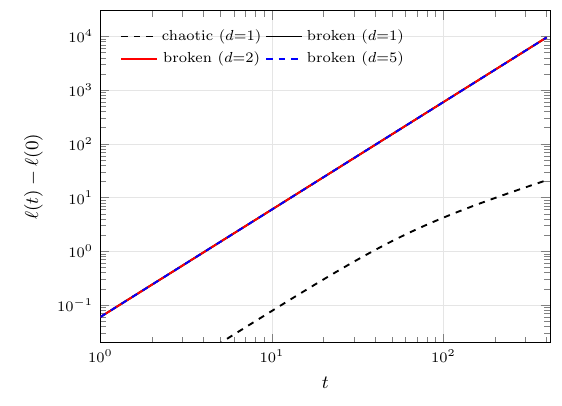}
  \caption{Real-time growth of the chord number (state complexity) at
  $\beta J=110$. Black dashed: chaotic phase ($\kappa=0.13$). Solid black,
  solid red and dashed blue: broken phase for $d=1,2,5$ evaluated at the
  \emph{same} $\kappa/\sqrt d=0.25$ ($\kappa=0.25$, $0.354$, $0.559$). The
  three broken-phase curves coincide exactly --- we draw the $d=5$ one dashed
  so that all three remain visible --- and follow $t^2$; the chaotic curve
  approaches a linear law at late times.}
  \label{fig:dyn}
\end{figure}

\subsection{Linear versus quadratic growth}

The result is shown in Fig.~\ref{fig:dyn}. In the broken phase the chord
number grows as
\begin{equation}
  \ell(t)-\ell(0) \simeq A\,t^2 ,
  \label{eq:t2}
\end{equation}
with a fitted exponent $2.00$ over two decades in $t$; this is the
quasi-integrable growth. In the chaotic phase the growth is much slower and
approaches a linear law,
\begin{equation}
  \ell(t)-\ell(0) \simeq B\,t ,
  \label{eq:t1}
\end{equation}
with a local exponent $\alpha(t)=d\ln(\ell-\ell_0)/d\ln t$ that decreases from
$1.8$ at $t=50$ to $1.07$ at $t=400$ and tends to $1$. The two behaviors are
separated by the same first-order transition as in the thermodynamics: at
fixed $\kappa$ the system is in the broken phase for
$d<(\kappa/\kappa_c(1))^2$ and in the chaotic phase otherwise.

The universality of the broken-phase dynamics is a direct consequence of the
mapping \eqref{eq:mapping}: the curves for $d=1,2,5$ in Fig.~\ref{fig:dyn}
are evaluated at the same $\kappa/\sqrt d=0.25$ and coincide exactly, and
curves at other values of $\kappa/\sqrt d$ differ from them only in overall
scale. This is the dynamical counterpart of the universal jump of
Table~\ref{tab:transition}.

\subsection{State versus operator complexity}

It is important to distinguish the state complexity \eqref{eq:CK}, which is
proportional to the chord number and grows as $t$ or $t^2$, from the Krylov
\emph{operator} complexity, which grows exponentially in the chaotic phase and
quadratically in the quasi-integrable phase \cite{Aguilar-Gutierrez:2026jjv,Balasubramanian:2022tpr,Rabinovici:2022beu}.
The two diagnostics carry the same information about the phase, but they are
not the same observable. In particular, the fact that the chord number grows
\emph{faster} in the quasi-integrable phase (as $t^2$ rather than $t$) does
not mean that the quasi-integrable phase scrambles faster: the operator
complexity, which is the standard chaos diagnostic, is exponential only in
the chaotic phase. Our results for the chord number are consistent with the
$d=1$ analysis of \cite{Aguilar-Gutierrez:2026jjv}, and the operator
complexity computed below exhibits the same discontinuity at $\kappa_c(d)$.

Guided by \cite{Aguilar-Gutierrez:2026jjv}, we compute the Lanczos
coefficients of the operator autocorrelation with the moment method. For a
light operator of dimension $\Delta$ the normalized autocorrelation at the
saddle is $G(t)=e^{-\Delta(\ell(t)-\ell(0))}$. In the limit $\Delta\to0$ its
moments are $\mu_0=1$, $\mu_{2n+1}=0$ and
\begin{equation}
  \mu_{2n} = -(-1)^n\Delta\,\ell^{(2n)}(0)\propto\Delta \qquad (n\ge1),
  \label{eq:moments}
\end{equation}
so that $b_1^2=\mu_2\propto\Delta$ while the higher coefficients are
$\Delta$-independent; this is the reason for the rescaling used below.
We evaluate $\ell^{(2n)}(0)$ from a Taylor expansion of the real-time
equations \eqref{eq:realtime} around $t=0$, working with exact arithmetic to
$200$ digits and keeping terms up to order $160$. The Lanczos coefficients
follow from the Hankel determinants
\begin{equation}
  b_1^2=\det H_1 ,\qquad
  b_n^2=\frac{\det H_n\,\det H_{n-2}}{(\det H_{n-1})^2} ,
  \qquad H_n=(\mu_{i+j})_{0\le i,j\le n} ,
  \label{eq:lanczos}
\end{equation}
and we use the rescaling $\tilde b_n=\sqrt{2\Delta}\,b_n/b_1$, which sets
$\tilde b_1=\sqrt{2\Delta}$ and leaves $\tilde b_{n\ge2}$ finite as
$\Delta\to0$.

\begin{figure}[t]
  \centering
  \includegraphics[width=0.95\textwidth]{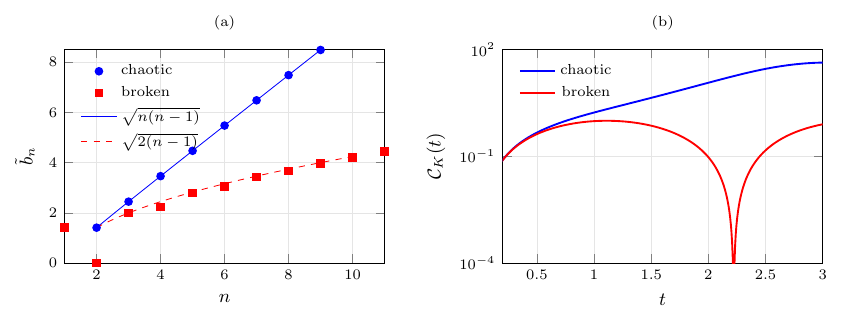}
  \caption{(a) Rescaled Lanczos coefficients $\tilde b_n$ of the operator
  autocorrelation at $\beta J=110$, for the chaotic phase ($\kappa=0.13$,
  blue) and the broken phase ($\kappa=0.30$, red). The thin lines are
  $\sqrt{n(n-1)}$ and $\sqrt{2(n-1)}$; the broken phase ($\ell\simeq267$) has
  $\tilde b_2\simeq6\times10^{-58}$. (b) The corresponding Krylov operator
  complexity $\mathcal C_K(t)$: the chaotic phase grows exponentially while
  the broken phase remains bounded. In the deep broken phase the coefficients
  become universal, $\tilde b_2\sim e^{-\ell/2}\to0$ and
  $\tilde b_{n\ge3}\simeq\sqrt{2(n-1)}$, so curves at different $d$ and
  $\kappa_{\rm eff}$ coincide: we show $d=1$ at $\kappa=0.30$
  ($\kappa_{\rm eff}=0.30$), and the $d=2$ and $d=5$ curves at $\kappa=0.35$
  and $0.55$ ($\kappa_{\rm eff}\simeq0.25$) fall on top of it.}
  \label{fig:opc}
\end{figure}

The result is shown in Fig.~\ref{fig:opc}. In the chaotic phase the rescaled
coefficients are
\begin{equation}
  \tilde b_n^2 = n(n-1)+\mathcal O\!\left(\frac{\kappa^4}{\tilde\nu^4}\right),
  \label{eq:bnchaotic}
\end{equation}
i.e.\ $\tilde b_n\propto n$, the hallmark of chaotic operator growth
\cite{Parker:2018yvk}.

In the broken phase the chain is instead
\begin{equation}
  \tilde b_2=\frac{2\nu}{\kappa}\,e^{g_z(\beta/2)/2}\ll1 ,
  \qquad
  \tilde b_n\simeq\sqrt{2(n-1)} \quad (n\ge3) ,
  \label{eq:bnbroken}
\end{equation}
i.e.\ $\tilde b_n\propto\sqrt n$ with an exponentially small first hop. Here
$g_z$ is the field of the \emph{active} integrable flavor, i.e.\ the one
carrying the large chord number (the spectators vanish), so that
$g_z(\beta/2)=-\ell+g_n(\beta/2)$, with $|g_n(\beta/2)|\ll\ell$ in the broken
phase, so that $\tilde b_2\sim e^{-\ell/2}$ (a saddle-point derivation is given
in Appendix~\ref{app:action}). At $\kappa=0.30$
one has $\ell\simeq267$ and $\tilde b_2\simeq6\times10^{-58}$, while at the
transition ($\ell\simeq70$) $\tilde b_2\sim10^{-15}$. In the broken phase the
coefficients become universal in the deep broken limit,
$\tilde b_2\sim e^{-\ell/2}\to0$ and $\tilde b_{n\ge3}\simeq\sqrt{2(n-1)}$,
independently of $d$ and of the precise value of $\kappa_{\rm eff}$; this is
why the $d=1,2,5$ curves coincide. In the chaotic phase the coefficients
depend only on $\kappa/\tilde\nu$, with
$\tilde\nu\equiv\nu e^{g_z(\beta/2)/2}$; the $d$ integrable fields add up to
$\kappa^2g_n$ at leading order in $\kappa$, so the effective source is
$d$-independent and the curves again coincide.

Equation~\eqref{eq:bnbroken} has a direct spectral interpretation. Writing the
moments of the normalized spectral function $\rho(\omega)$ as
$\mu_{2n}=\int d\omega\,\omega^{2n}\rho(\omega)$, one has
\begin{equation}
  \tilde b_2^2=2\Delta\left(\frac{\mu_4}{\mu_2^2}-1\right)
  \equiv 2\Delta\,\kappa_{\rm ex},
  \label{eq:kurtosis}
\end{equation}
i.e.\ $\tilde b_2$ is the excess kurtosis of the spectral function:
$\kappa_{\rm ex}\simeq2$ for the Gaussian-like chaotic spectrum and
$\kappa_{\rm ex}\to0$ for the $\delta$-like quasi-integrable one. At the
transition $\tilde b_2^2$ drops from $1.997$ to $2.97\times10^{-29}$, a change
of nearly thirty orders of magnitude.

Evolving the Krylov wavefunction with these coefficients gives the operator
complexity $\mathcal C_K(t)=\sum_n n|\varphi_n(t)|^2$ shown in panel (b) of
Fig.~\ref{fig:opc}. Both phases exhibit the universal early-time growth
$\mathcal C_K\simeq 2t^2$; in the chaotic phase this crosses over to an
exponential, $\mathcal C_K\sim e^{\lambda_K t}$, with a fitted
$\lambda_K\simeq1.84$ at $n_c=80$ that increases toward the value
$\lambda_K=2$ expected for $\tilde b_n\propto n$ \cite{Parker:2018yvk} as
$n_c$ grows ($1.89$ at $n_c=100$), while in the
broken phase the exponentially small $\tilde b_2$ effectively disconnects the
chain, and $\mathcal C_K$ remains bounded, oscillating with revivals (the dip
near $t\simeq2.2$ in Fig.~\ref{fig:opc}) rather than growing. The transition at
$\kappa_c(d)$ therefore appears as a discontinuity in $\tilde b_2$ and in the
late-time growth of the operator complexity, consistent with the $d=1$
analysis of \cite{Aguilar-Gutierrez:2026jjv}. The Krylov exponent is expected to
be related to the chaos exponent $\lambda_L=2\pi v/\beta$ of the double-scaled
SYK model \cite{Lin:2023}, with $\beta J=\pi v/\cos(\pi v/2)$; our $\lambda_K$ is
obtained in the rescaled Krylov chain. That rescaling is a change of time
unit: $\tilde b_n=\sqrt{2\Delta}\,b_n/b_1$ amounts to measuring the Krylov time
in units of $b_1/\sqrt{2\Delta}$, so restoring the physical normalization
multiplies any fitted growth rate by the same factor. A quantitative
comparison with $\lambda_L$ therefore requires fixing the normalization of
$\Delta$ and of the map from chord number to state complexity, which we leave
for future work. The coefficients are stable to six digits under changes of the working
precision ($160$ to $240$ digits) and of the moment truncation ($n_{\rm
max}=80$ to $160$); the chain is truncated at $n_c=80$, and the exponential
window $0.5\lesssim t\lesssim3$ shown in the figure is free of truncation
artifacts for the parameters used.

\section{Conclusions}
\label{sec:conclusions}

We have revisited the fate of the BBJM first-order chaos--integrability
transition in the presence of $d$ non-commuting integrable flavors. Our main
results are:

\begin{enumerate}
  \item In the saddle-point (annealed) approximation, the color-resolved chord
  path integral admits a flavor-symmetry-breaking saddle in which a single
  integrable flavor carries the quasi-integrable phase. Up to power-law-small spectator corrections, its active pair
  obeys the $d=1$ saddle equations with $\kappa\to\kappa/\sqrt d$, and its
  action equals that of the normalized BBJM saddle at $\kappa/\sqrt d$ to
  $\mathcal O\!\left((\beta J)^{-2}\right)$; among the saddles examined here, it has the lowest
  on-shell action at large $\kappa$ for every $d<\infty$.
  \item The transition therefore survives, at
  $\kappa_c(d)\simeq\sqrt d\,\kappa_c(1)$. It leaves the physical window
  $\kappa\le1$ only for $d>d_c(\beta J)=1/\kappa_c(1)^2$, which we determined
  numerically; $d_c\simeq33.5$ at $\beta J=110$, $d_c\simeq0.30\,\beta J$
  in the range $30\le\beta J\le240$, and $d_c\simeq4.9$ at the critical
  temperature $\beta J_c\simeq13.5$.
  \item The jump of the chord number at the transition is universal,
  $\ell:7\to70$, independent of $d$. The dynamics follows the same pattern:
  $t^2$ growth of the chord number (state complexity) in the broken phase and
  linear growth in the chaotic phase; the Lanczos coefficients of the operator
  autocorrelation change from $\tilde b_n\propto n$ to
  $\tilde b_n\propto\sqrt n$ for the BBJM family ($l_{zz}=0$), with an exponentially small $\tilde b_2$ in the
  broken phase (for the general two-chord deformation with $l_{zz}>0$ the broken
  branch instead shows $\tilde b_n^2\propto n(n-1)$, see Sec.~\ref{sec:twoc}); and the
  corresponding operator complexity is exponential in
  the chaotic phase and bounded in the broken phase.
  \item The color-averaged weight $\bar q(d)$ of Gao, Lin and Peng is the
  correct leading large-$d$ description and a useful benchmark; at finite $d$
  it enforces flavor symmetry and misses the broken saddle. The two
  descriptions agree for $d\gg1$ and diverge at $d=2$: the average predicts
  melting, the color-resolved saddle-point model does not. This divergence is
  the finite-$d$ physics, not a failure of either description. The same
  remark applies to the symmetric two-chord
  deformation $l_{zz}\neq0$ of \cite{Berkooz:2024ofm}: it is not the
  deformation that the microscopic multi-color model realizes.
\end{enumerate}

It is worth separating what is robust in these results from what is assumed.
The reduction \eqref{eq:mapping}, the approximate $\sqrt d$ law
\eqref{eq:kc}, the melting line \eqref{eq:dc2} and the universal jump are
consequences of the saddle-point equations and of the on-shell formula
\eqref{eq:onshell}, which at $d=1$ reproduces the known BBJM values
$-2\beta J+\pi^2/2$ and $-\beta^2/2$; the dynamical results follow from the
same saddles. The assumptions are: (i) the saddle-point (annealed) evaluation
of the ensemble-averaged free energy; (ii) dominance of the broken saddle in
the sense of the lowest on-shell action among the saddles examined
(Tables~\ref{tab:transition} and~\ref{tab:kactive}), not a global thimble
decomposition; (iii) neglect of finite-$N$ tunneling between the $d$ broken
vacua, which is of order $e^{-cN}$ and does not affect the large-$N$ saddle.
The fluctuation Hessian of the broken saddle is indefinite in its homogeneous
sector, as it is for the DSSYK saddle itself; the Morse index affects the
phase of each saddle and not its magnitude at leading order in $\lambda$, and
the induced one-loop shift of $\kappa_c$ is of order $10^{-3}$, while the
one-loop \emph{prefactor} is left open (App.~\ref{app:hess}). Throughout $d$
is held fixed as $N\to\infty$, so the chord rules are exact in this limit and
$d_c\simeq0.30\,\beta J\ll N$ whenever $\beta J\ll N$.

Let us reconcile this picture with GLM. At $\kappa=1$ our model reduces to the
pure $d$-commuting SYK and the broken saddle is its low-temperature
flavor-broken phase. GLM distinguish two limits \cite{Gao:2024lve}: for $q\to1$
and $d<9$ they argue that the crossover at $T_c$ has thermodynamic meaning,
with self-averaging only for $T\gtrsim T_c$ and a new phase below it,
whereas for $q=0$ the crossover at $T_{\rm GLM}=J/(2d)$ is only spectral. Our chord description is the
$\lambda\to0$ ($q\to1$) semiclassical limit, so the $q\to1$ case is the
relevant comparison. Our $T_{\rm re}(d)\simeq0.30\,J/d$ decreases with $d$ as
GLM's $T_c$ does and lies a factor of about $1.7$ below the $q=0$ value
$J/(2d)$, which we use only as the simplest $1/d$ reference scale. The
comparison is therefore qualitative rather than quantitative: GLM's one-loop
determination covers $d<9$, whereas our transition survives up to $d\simeq33$
at $\beta J=110$. The two pictures are complementary --- GLM identify the
temperature above which the $d$-commuting model behaves chaotically, while we
identify the low-temperature phase that replaces the chaotic one.

Several directions are worth pursuing. First, it would be interesting to compute the OTOC across the flavor-breaking
transition, and to study the operator complexity with the full Liouvillian in
the chord basis, to confirm that they exhibit the same discontinuity as the
Lanczos coefficients. Second, the finite-$N$ fate of the broken saddle deserves a careful
study. At finite $N$ the permutation symmetry is restored by tunneling
between the $d$ broken vacua; the tunneling amplitude is of order
$e^{-\Delta S/\lambda}$ with $\Delta S$ the barrier action, and since
$\lambda=2p^2/N\to0$ in the double-scaling limit the restoration is
non-perturbative in $1/N$. Estimating $\Delta S$ would provide a sharp
criterion for the observability of the broken phase. Third, it would be
natural to look for a holographic interpretation of the broken saddle, perhaps
in terms of a condensate of a single boundary flavor in the sine-dilaton
gravity dual of DSSYK \cite{Blommaert:2024ydx}. Finally, the same mechanism may apply to other
multi-flavor chord models, such as the Parisi hypercube model with several
species of fluxes.

\begin{acknowledgments}

We thank our colleagues at Shanghai University for useful discussions.

\end{acknowledgments}

\appendix

\section{Derivation of the multi-field action and its on-shell value}
\label{app:action}

In this appendix we derive \eqref{eq:S} and \eqref{eq:onshell}. The starting
point is the chord diagram expansion of the moments,
\begin{equation}
  Z(\beta) = \sum_{k=0}^{\infty}\frac{(-\beta)^k}{k!}
  \langle \Tr \hat H^k\rangle ,
\end{equation}
where each moment is a sum over colored chord diagrams with $n_a$ chords of
species $a$. A diagram with $n_a$ chords of species $a$ carries the weight
\begin{equation}
  \prod_{a=0}^{d}\left(\kappa_a^2J^2\right)^{n_a}
  q^{\sum_{a<b}\M_{ab}\,C_{ab}} ,
\end{equation}
where $C_{ab}$ is the number of crossings between chords of species $a$ and
$b$, and the crossing weight is $q^{\M_{ab}}$. Coarse graining the diagrams as
in \cite{Berkooz:2024ofm} (generate, split, reorder, cross) and introducing the
chord densities $n_a(\tau_1,\tau_2)$ gives \eqref{eq:S}, with the same
normalization as in the two-chord case. The only difference is the
intersection matrix $\M_{ab}$ in the crossing term and the couplings
$\kappa_a^2$ in the entropy term; the steps of the derivation are identical to
those of \cite{Berkooz:2024ofm}.

Varying \eqref{eq:S} with respect to $n_a$ gives
\begin{equation}
  \sum_{b=0}^{d}\M_{ab}\int_{\tau_1}^{\tau_2}d\tau_3
  \int_{\tau_2}^{\tau_1}d\tau_4\, n_b(\tau_3,\tau_4)
  + \ln\frac{n_a(\tau_1,\tau_2)}{\kappa_a^2J^2} = 0 ,
\end{equation}
which, using \eqref{eq:g_def}, is
\begin{equation}
  n_a = \kappa_a^2J^2\exp\left[\sum_{b=0}^{d}\M_{ab}\,g_b\right] .
\end{equation}
Combining with $n_a=-\frac12\partial_1\partial_2 g_a$ gives
\eqref{eq:eom}. To obtain the on-shell action, we use the identity
\begin{equation}
  \frac14\int\!\!\int\!\!\int\!\!\int
  \sum_{a,b}\M_{ab}n_a n_b
  = -\frac14\int\!\!\int\sum_{a,b}\M_{ab}\,n_a g_b ,
\end{equation}
and the saddle-point relation
$\ln(n_a/\kappa_a^2J^2)=\sum_b\M_{ab}g_b$. Substituting into \eqref{eq:S}
and using $n_a=-\frac12\partial_1\partial_2 g_a$ gives
\begin{equation}
  S_{\rm on\text{-}shell}
  = -\frac18\int\!\!\int\sum_{a,b}\M_{ab}\,g_a\,
  \partial_1\partial_2 g_b
  -\frac12\int\!\!\int\sum_a\kappa_a^2J^2\,
  e^{\sum_b\M_{ab}g_b} .
\end{equation}
For a saddle that depends only on $\tau=\tau_1-\tau_2$, the double integral
becomes $\beta\int_0^\beta d\tau$ and
$\partial_1\partial_2\to-\partial_\tau^2$, which yields \eqref{eq:onshell}.
For $d=1$ one recovers \eqref{eq:BBJM_onshell} after using the equations of
motion. The two expressions are not term-by-term identical, but they agree
identically on shell. Written in terms of the fields, their difference is a
total derivative,
\begin{equation}
\begin{split}
S_{\rm on\text{-}shell}-\big[\text{\eqref{eq:BBJM_onshell} at }d=1\big]
&= \frac{\beta}{8J^2}\int_0^\beta d\tau\,
\partial_\tau\big(g_z g_n'-g_n g_z'\big) \\
&= \frac{\beta}{8J^2}\Big[g_z g_n'-g_n g_z'\Big]_0^\beta ,
\end{split}
\label{eq:totalderiv}
\end{equation}
where we used $e^{g_n+g_z}=g_n''/(2J^2\nu^2)$ and
$e^{g_n}=g_z''/(2J^2\kappa^2)$. The boundary term vanishes because
$g_n=g_z=0$ at $\tau=0$ and at $\tau=\beta$, so the requirement that the
general formula reduce to the two-field one is satisfied identically rather
than only approximately. We have checked \eqref{eq:totalderiv} numerically at
$\kappa=0$, $0.13$ and $0.173$, where the two forms agree to four digits.

\paragraph*{Saddle-point derivation of $\tilde b_2$.}
In the broken phase the $n$-chords are frozen, $g_n\simeq0$, and the active
flavor obeys $g_z(\tau)=-\kappa^2\tau(\tau-\beta)$, so that
$g_z(\beta/2)=-\kappa^2\beta^2/4\equiv-\ell$. On the real-time contour we write
$g_z(t)=-\ell+\kappa^2t^2+\cdots$; the $n$-field equation
$g_n''=2\nu^2e^{g_n+g_z}$ then gives
\begin{equation}
  g_n(t)=\nu^2e^{-\ell}t^2+\frac{\nu^2\kappa^2}{6}e^{-\ell}t^4+\cdots .
\end{equation}
With $\ell(t)=g_n(t)+g_z(t)$ and
$\mu_{2n}=-(-1)^n\Delta\,\ell^{(2n)}(0)$ one finds
\begin{equation}
  \mu_2=2\Delta\left(\kappa^2+\nu^2e^{-\ell}\right),\qquad
  \mu_4=4\Delta\nu^2\kappa^2e^{-\ell},
\end{equation}
hence
\begin{equation}
  b_2^2=\frac{\mu_4}{\mu_2}-\mu_2
  \simeq 2\nu^2e^{-\ell}-2\Delta\kappa^2
  \;\xrightarrow[\Delta\kappa^2\ll\nu^2e^{-\ell}]{}\;2\nu^2e^{-\ell}.
\end{equation}
After the rescaling $\tilde b_n=\sqrt{2\Delta}\,b_n/b_1$ this yields
\begin{equation}
  \tilde b_2\propto\frac{\nu}{\kappa}\,e^{-\ell/2}.
\end{equation}
The exponential scaling and the $\nu/\kappa$ dependence are convention
independent; the $O(1)$ coefficient depends on the normalization of $\Delta$
and $b_1$ and equals $2$ in the convention used in the main text. The
numerical values at $\kappa=0.18$, $0.20$, $0.25$ and $0.30$ agree with
Eq.~\eqref{eq:bnbroken} to $22\%$, $12\%$, $4\%$ and $2\%$, respectively.

\paragraph*{Bilinear fluctuation Hessian.}\label{app:hess}
In the chord-density variables the second variation of \eqref{eq:S} is
\begin{equation}
  \mathcal H = \tfrac12\,\M\otimes K + \mathrm{diag}\Big(\frac{1}{n_a}\Big),
  \qquad
  (K\delta n)(\tau_1,\tau_2)=\int_{\tau_1}^{\tau_2}\!\!d\tau_3
  \int_{\tau_2}^{\tau_1}\!\!d\tau_4\,\delta n(\tau_3,\tau_4),
  \label{eq:hessn}
\end{equation}
with $K$ the crossing operator and $1/n_a$ the second variation of the
entropy. This form is too degenerate to serve as a stability criterion. The
entries of $K$ depend on $(\tau_1,\tau_2)$ only through the ordered pair
$(\min,\max)$, so on an $N\times N$ grid $K$ has rank at most
$\binom{N+1}{2}$ and the number of zero modes is of order $N^2$; moreover
$n_a$ vanishes identically in the frozen and spectator sectors, so $1/n_a$
must be regulated, and the resulting spectrum spans more than ten orders of
magnitude, with $\lambda_{\min}/\lambda_{\max}$ of order $10^{-11}$ and of
either sign. The sign of the near-zero directions is therefore not resolved at
accessible grid sizes, and the one-loop prefactor, which would require a
Faddeev--Popov treatment of these directions, is left open. The selection of
the dominant saddle is made by the on-shell action, as discussed in
Sec.~\ref{sec:stability}, and does not rely on this analysis.

\section{Numerical methods and checks}
\label{app:numerics}

We solve the saddle-point equations \eqref{eq:eom_tau} on
$\tau\in[0,\beta/2]$ with the boundary conditions $g_a(0)=0$,
$g_a'(\beta/2)=0$ (the solutions are symmetric about $\tau=\beta/2$). The
interval is discretized on $N+1$ Chebyshev--Gauss--Lobatto points,
\begin{equation}
  \tau_j = \frac{\beta}{4}\left(1-\cos\frac{\pi j}{N}\right) ,
  \qquad j=0,\dots,N ,
\end{equation}
and the second derivative is represented by the Chebyshev differentiation
matrix. The nonlinear system is solved by Newton's method with a
line search; the Jacobian is computed analytically. We use $N=120$--$160$ and
require a residual below $10^{-9}$. The on-shell action is evaluated by
interpolating the integrand onto a fine uniform grid and using the trapezoidal
rule. For the numerical evaluation we use the equivalent form
\begin{equation}
  S_{\rm on-shell}=\frac{\beta}{4}\int_0^\beta d\tau\,
  \sum_a\kappa_a^2\,e^{G_a}\left(G_a-2\right) ,
  \qquad G_a\equiv\sum_b \M_{ab}g_b ,
  \label{eq:Snum}
\end{equation}
obtained from the equations of motion, which avoids differentiating the
solution twice.

The convergence of the discretization was checked explicitly (Table~\ref{tab:conv}). The fields
converge to $10^{-6}$ at $N=160$. The absolute value of the on-shell action
carries a residual discretization error of about $0.1$ ($5\times10^{-4}$
relative) at $N=160$, as estimated by comparing $N=120,160,200$. This error is
almost entirely additive: the \emph{differences} between saddles evaluated on
the same grid are stable to $3\times10^{-4}$ over the same range of $N$,
because the discretization errors cancel. All comparisons of saddles in the
text (the mapping checks, the transition points, and the one-loop shift
estimate) use such differences, so they are not affected by the absolute
normalization error.

\begin{table}[H]
  \centering
  \begin{tabular}{c|cc|cc|c}
    \hline
    $N$ & $S_{\rm ch}$ ($d=1$) & $\ell_{\rm ch}$ & $S_{\rm br}$ ($d=2$)
        & $\ell_{\rm br}$ & $S_{\rm br}-S_{\rm BBJM}$ \\
    \hline
    80  & $-215.909$ & 7.14958 & $-763.948$ & 374.593 & $-3.4\times10^{-4}$ \\
    120 & $-215.507$ & 7.14958 & $-763.577$ & 374.595 & $-3.4\times10^{-4}$ \\
    160 & $-215.367$ & 7.14958 & $-763.446$ & 374.595 & $-3.4\times10^{-4}$ \\
    200 & $-215.301$ & 7.14958 & $-763.385$ & 374.595 & $-3.4\times10^{-4}$ \\
    \hline
  \end{tabular}
  \caption{Convergence of the discretization for the chaotic saddle at $d=1$
  ($\kappa=0.13$) and the broken saddle at $d=2$ ($\kappa=0.5$). The absolute
  actions drift with $N$ (an additive discretization error), while the chord
  numbers and the difference between the full broken saddle and the $d=1$
  saddle at $\kappa_{\rm eff}=0.5/\sqrt2$ are stable. The latter is the
  quantity used in the mapping checks.}
  \label{tab:conv}
\end{table}

The two competing saddles are found by continuation. The chaotic branch is
tracked from $\kappa=0$, where the solution is \eqref{eq:chaotic_g}, upward in
$\kappa$. The broken branch is tracked from $\kappa=0.99$, where
$g_1=(\kappa^2/d)\,\tau(\tau-\beta)$ and the other fields vanish, downward in
$\kappa$. The full $(d+1)$-field system is used for $d=2,3,5$; for
$d=10,20,30$ we use the exact mapping \eqref{eq:mapping}, which we have
verified against the full system at $d=2,3,5$.

As a check of the whole pipeline, for $d=1$ we reproduce the BBJM results:
at $\beta J=110$ the chaotic saddle has $\ell=7.148$ and
$S=-215.36$, to be compared with the analytic value
$-2\beta J+\pi^2/2=-215.07$; the integrable saddle at $\kappa=1$ has
$\ell=3025$ and $S=-6050=-\beta^2/2$, as required by the Gaussian spectrum of
the commuting-SYK model. The transition point $\kappa_c(1)=0.173$ is
consistent with the low-temperature formula $\kappa_c\simeq2/\sqrt{\beta J}$.

The code used for the numerical results in this paper is available from the
authors upon reasonable request.

\bibliographystyle{apsrev4-2}
\bibliography{refs}

\end{document}